\documentclass[10pt,aps,prl,twocolumn,superscriptaddress,floatfix,showpacs,longbibliography]{revtex4-1}

\usepackage{graphicx}
\usepackage{amsfonts}
\usepackage{amssymb}
\usepackage{amsmath}
\usepackage{txfonts}
\usepackage{lipsum}
\usepackage{color}
\usepackage{wasysym}
\usepackage{hyperref}
\usepackage{bbold}
\usepackage{etoolbox} 
\usepackage[normalem]{ulem}
\usepackage{mathtools}
\usepackage{multirow}
\usepackage{hhline}
\usepackage{mathrsfs} 
\usepackage{soul}

\newcommand{\av}[1]{\ensuremath{\left\langle #1 \right\rangle}}

\usepackage{letltxmacro}
\LetLtxMacro{\oldsqrt}{\sqrt}
\renewcommand{\sqrt}[2][\mkern8mu]{\mkern-6mu\mathop{}\oldsqrt[#1]{#2}}

\usepackage{url}
\definecolor{indigo(dye)}{rgb}{0.0, 0.25, 0.42}
\usepackage{placeins}

\begin{document}

\title{Dynamically induced doublon repulsion in the Fermi-Hubbard model \\
probed by a single-particle density of states}

\author{V. N. Valmispild}
\affiliation{I. Institute of Theoretical Physics, University of Hamburg, Jungiusstrasse 9, 20355 Hamburg, Germany}
\affiliation{The Hamburg Centre for Ultrafast Imaging, Luruper Chaussee 149, 22761 Hamburg, Germany}
\affiliation{European XFEL, Holzkoppel 4, 22869 Schenefeld, Germany}

\author{C. Dutreix}
\affiliation{Univ. Bordeaux, CNRS, LOMA, UMR 5798, F-33400 Talence, France}

\author{M. Eckstein}
\affiliation{Department of Physics, University of Erlangen-Nuremberg, 91058 Erlangen, Germany}

\author{M. I. Katsnelson}
\affiliation{Radboud University, Institute for Molecules and Materials, 6525AJ Nijmegen, The Netherlands}

\author{A. I. Lichtenstein}
\affiliation{I. Institute of Theoretical Physics, University of Hamburg, Jungiusstrasse 9, 20355 Hamburg, Germany}
\affiliation{The Hamburg Centre for Ultrafast Imaging, Luruper Chaussee 149, 22761 Hamburg, Germany}
\affiliation{European XFEL, Holzkoppel 4, 22869 Schenefeld, Germany}

\author{E. A. Stepanov}
\affiliation{I. Institute of Theoretical Physics, University of Hamburg, Jungiusstrasse 9, 20355 Hamburg, Germany}

\begin{abstract}
We investigate the possibility to control dynamically the interactions between repulsively bound pairs of fermions (doublons) in correlated systems with off-resonant ac fields.
We introduce an effective Hamiltonian that describes the physics of doublons up to the second-order in the high-frequency limit.
It unveils that the doublon interaction, which is attractive in equilibrium, can be completely suppressed and then switched to repulsive by varying the power of the ac field.
We show that the signature of the dynamical repulsion between doublons can be found in the single-fermion density of states averaged in time.
Our results are further supported by nonequilibrium dynamical mean-field theory simulations for the half-filled Fermi-Hubbard model.
\end{abstract}

\maketitle

The ability to control matter by strong laser pulses has always intrigued researchers in many areas of physics. Recent development of femtosecond laser sources allows to perform experiments on a time-scale of the dominant microscopic interactions in materials. These experiments offer an
outstanding possibility to selectively excite different collective modes, which has lead to intriguing results, such as light-induced magnetism~\cite{PhysRevLett.76.4250, kimel2005ultrafast, RevModPhys.82.2731,  koopmans2010explaining, PhysRevLett.107.076601, satoh2012directional, schellekens2014ultrafast, stupakiewicz2017ultrafast}, superconductivity~\cite{Fausti189, PhysRevB.95.134508}, and topological states of matter~\cite{PhysRevB.79.081406, lindner2011floquet, PhysRevLett.112.156801, PhysRevB.93.241404, PhysRevLett.118.157201, mciver2020light}. 
Investigation of field-driven effects in fermionic systems is also motivated by inspiring results of ultracold atom physics~\cite{doi:10.1080/00018730701223200, RevModPhys.80.885, RevModPhys.89.011004}, where the effect of the applied perturbation can be mimicked by a modulation of the position (shaking) of the lattice~\cite{PhysRevLett.99.220403, PhysRevLett.108.225304, PhysRevLett.109.145301, rechtsman2013photonic, jotzu2014experimental}, or through the engineering of photon-assisted hopping amplitudes~\cite{Goldman_2014, RevModPhys.89.011004}.

Among different light-induced collective excitations, a big attention of the experimental~\cite{exp_3, exp_1, exp_4, Wall_1} and theoretical~\cite{PhysRevB.85.205127, Eckstein_1, Eckstein_3, RevModPhys.86.779, Eckstein_2, Wang_1, Rausch_1, PhysRevLett.120.166401, PhysRevB.99.205108} condensed matter physics is devoted to repulsively bound pairs of fermions that occupy the same lattice site. 
Effects related to these composite bosonic objects, known as doublons, are also actively discussed in the context of cold atoms~\cite{Bloch_2, Winkler_1, RevModPhys.80.885, Trefzger_1, doi:10.1080/00018730701223200, PhysRevA.84.021607, Cheuk1260}. 
Interestingly, the concept of doublons has been introduced as early as in 1930th, within the so called ``polar model''~\cite{doi:10.1098/rspa.1934.0089} (for more modern presentation see~\cite{Vonsovsky_1979_1, Vonsovsky_1979_2}).
Theoretically, dynamics of doublons can be studied in the Mott-insulating regime of the Fermi-Hubbard model, where these bosonic quasiparticles have an exponentially large lifetime due to a strong repulsive on-site Coulomb interaction~\cite{PhysRevLett.101.265301, PhysRevLett.107.145303, PhysRevB.84.035122}. In this case, a fingerprint of doublon excitations is contained in a fermion density of states (DOS), where states related to doubly-occupied lattice sites form upper and lower Hubbard sub-bands. The latter can be efficiently observed in the (inverse) photoemission spectroscopy experiments~\cite{Neddermeyer_1,COURTHS198453,RevModPhys.75.473}. 

One of the main interests in doublons is associated with the effect of Bose-Einstein condensation (BEC)~\cite{Winkler_1, Jochim2101, PhysRevLett.91.250401, greiner2003emergence, PhysRevA.76.033606} and phase transition from an insulating to a superfluid state~\cite{greiner2002quantum, doi:10.1143/JPSJ.80.084607}. In equilibrium, the interaction between doublons is attractive~\cite{CHAO1977163, Chao_1977, PhysRevB.37.9753, spalek2007tj}. 
Since the pioneering work by Valatin and Butler~\cite{valatin1958collective}, it is known that the Bose gas with attractive interactions has a tendency to a phase separation (see also~\cite{PhysRevLett.76.2670, PhysRevA.54.661, PhysRevA.54.3151, PITAEVSKII199614}). However, the BEC can be achieved introducing a short-range repulsion in the system~\cite{refId0, PhysRevB.37.5893, PhysRevB.65.195103}. Thus, a dynamical control of the doublon interaction can completely change properties of the system and may allow for a precise control of these effects.
For example, it has been shown that the local Coulomb interaction in the Fermi-Hubbard model can be effectively switched from repulsive to attractive applying a periodic perturbation. This can be achieved creating of a population inversion in electronic bands through the sign change of the hopping amplitude~\cite{PhysRevLett.105.220405, PhysRevLett.106.236401}, or by a properly chosen pulse shape~\cite{PhysRevB.85.155124}. 
Later, this result has been used to modify an effective interaction between doublons, which resulted in the change of the superfluidity pairing from $s$-wave to $\eta$-pairing~\cite{PhysRevB.94.174503, PhysRevLett.122.077002, PhysRevB.101.161101, li2019longrange}.

Here, we propose a novel nonequilibrium mechanism to switch the doublon interaction from attractive to repulsive with an ac field.
We derive an effective time-independent Hamiltonian that describes the doublon physics up to second order in the high-frequency limit of the field.
It reveals that, in contrast to the works mentioned above, the repulsion between doublons is induced without population inversion and also for considerably smaller powers of the field.
Importantly, we argue that this interaction switch can be detected experimentally in a simple way via the single-fermion density of states (DOS) averaged in time.
Such an observable does not require the use of time-resolved techniques and, therefore, could be routinely measured in experiments.
We further support these findings numerically with nonequilibrium dynamical mean-field theory (DMFT) simulations~\cite{RevModPhys.68.13}.
Our result suggests that for detection of Floquet interactions in solids the doublon channel may be a good alternative to the spin exchange interaction.
The control of the spin exchange has been demonstrated in cold atoms~\cite{gorg2018enhancement}, but a measurement in the solid would require the use of a much more complicated time-resolved resonant inelastic X-ray scattering technique. 
On the contrary, the DOS can be measured even when the system is highly excited by photo-doping, which suppresses spin correlations. 

{\it High-frequency doublon Hamiltonian ---}
We consider the time-periodic Fermi-Hubbard Hamiltonian on a square lattice
\begin{align}
H &= \sum_{\av{ij},\,\sigma}t^{\phantom{\dagger}}_{ij}(\tau)c^{\dagger}_{i\sigma}c^{\phantom{\dagger}}_{j\sigma}
+ U \sum_{i} n^{\phantom{\dagger}}_{i\uparrow}n^{\phantom{\dagger}}_{i\downarrow}.
\label{eq:Hamiltonian1}
\end{align}
Operator $c_{i\sigma}$ annihilates an electron on site $i$ with spin $\sigma$, $n^{\phantom{\dagger}}_{i\sigma}=c^{\dagger}_{i\sigma} c^{\phantom{\dagger}}_{i\sigma}$ is the electron number operator, and $U$ the repulsive on-site Coulomb potential.
The time periodicity arises from a uniform ac perturbation of frequency $\Omega$, directed along the square lattice diagonal ${\bf e}=\{1,1\}$, as implemented, for example, with an electric field driving the electrons of a material, or a shaken lattice of cold atoms~\cite{PhysRevLett.99.220403, jotzu2014experimental, rechtsman2013photonic, PhysRevLett.108.225304, PhysRevLett.109.145301}.
The field is incorporated via a vector potential ${\bf A}(\tau)= A {\bf e}\cos(\tau)$, where time $\tau$ is given in units of $\Omega^{-1}$.
The hopping amplitude between nearest-neighbors $\av{ij}$ then accumulates a Peierls phase and satisfies $t_{ij}(\tau) = t e^{-i{\bf A}(\tau)\cdot{\bf R}_{ij}}$, where ${\bf R}_{ij}$ is the unit real-space vector between neighboring sites.

In equilibrium, the Fermi-Hubbard Hamiltonian~\eqref{eq:Hamiltonian1} maps via a Schrieffer-Wolff transformation onto an effective model that describes the low-energy physics of doublons in the limit $U\gg t$~\cite{CHAO1977163, Chao_1977, PhysRevB.37.9753, spalek2007tj}.
The presence of an external time-dependent field complicates this task.
So, we first perform a Magnus-like expansion in the spirit of Refs.~\cite{ITIN2014822, PhysRevLett.115.075301, PhysRevLett.116.125301, PhysRevB.93.241404, PhysRevB.95.024306, PhysRevLett.118.157201, PhysRevB.94.174503, PhysRevB.101.161101}.
This allows us to derive an effective time-independent Hamiltonian that captures the field renormalization of the fermion hopping and interaction up to the second order in the high-frequency limit $U\ll \Omega$
\begin{align}
H'(A) &= \sum_{\av{ij},\,\sigma}t'(A)c^{\dagger}_{i\sigma}c^{\phantom{\dagger}}_{j\sigma}
+ U'(A) \sum_{i} n^{\phantom{\dagger}}_{i\uparrow}n^{\phantom{\dagger}}_{i\downarrow} \notag\\
&+ \sum_{\av{ij}} \left(J'(A)\,d^{\dagger}_{i}d^{\phantom{*}}_{j} + \frac12V'(A)\,n_{i}n_{j}
+ \frac12{\cal I}'(A)\,{\bf S}_{i}^{\phantom{*}}{\bf S}_{j}^{\phantom{*}}\right).
\label{eq:Hp}
\end{align}
Here, the electron hopping $t'(A)$ and the local Coulomb interaction $U'(A)$ explicitly depend on the amplitude of the applied field through $A$. In addition, the high-frequency field induces purely nonequilibrium two-particle processes described by the nonlocal Coulomb potential $V'(A)$, exchange interaction strength ${\cal I}'(A)$, and doublon hopping amplitude $J'(A)$, for which we introduced the doublon operator $d_{j} = c_{j\downarrow}c_{j\uparrow}$ that annihilates a pair of fermions on site $j$. Their explicit expressions are in the Supplemental Material~\cite{SM}. Local charge and spin densities are defined as $n_{i}=\sum_{\sigma}n_{i\sigma}$ and ${\bf S}^{\phantom{\dagger}}_{i} = \frac12\sum_{\sigma,\sigma'}c^{\dagger}_{i\sigma} \boldsymbol{\sigma}_{\sigma\sigma'}c^{\phantom{\dagger}}_{i\sigma'}$, respectively. $\boldsymbol{\sigma}=\{\sigma^{x}, \sigma^{y}, \sigma^{z}\}$ is a vector of Pauli matrices.

\begin{figure}[t!]
\includegraphics[width=0.95\linewidth]{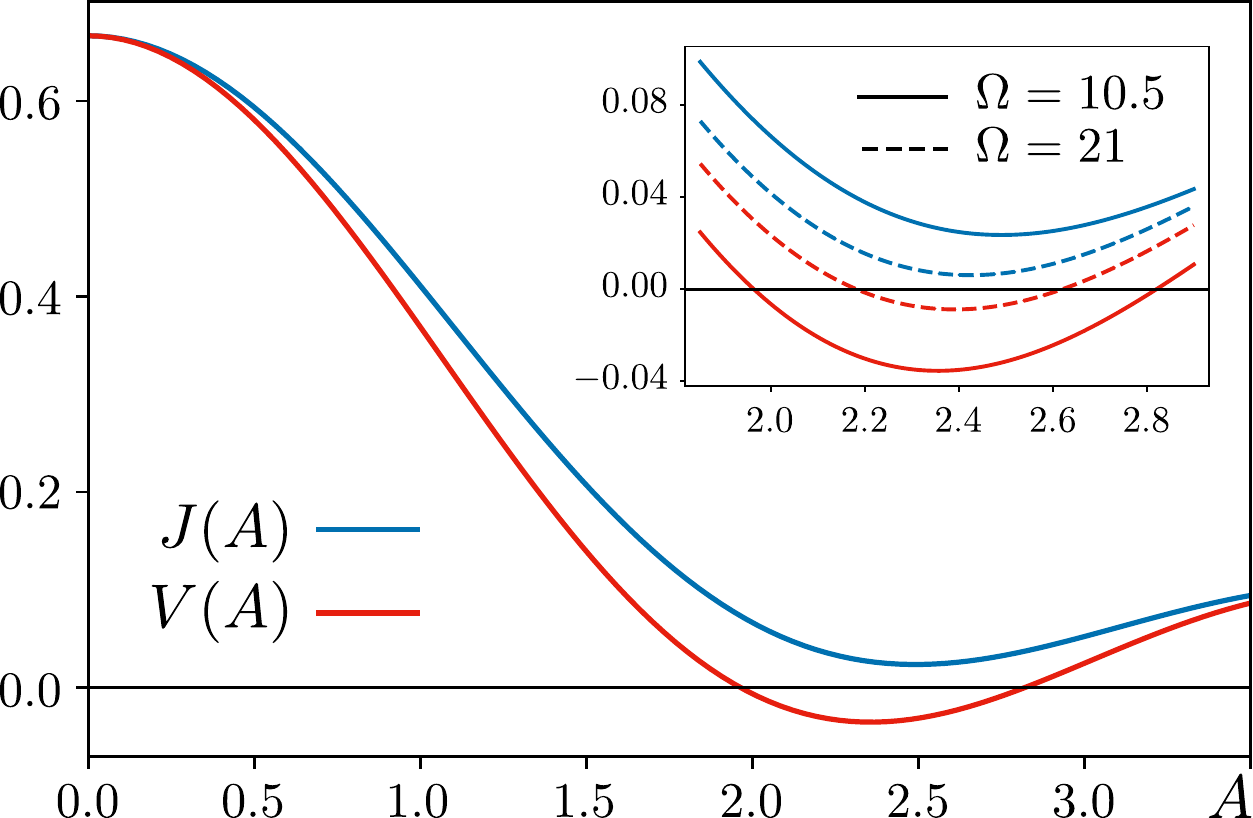}
\caption{\label{fig:TV} Doublon parameters of the effective Hamiltonian~\eqref{eq:EffHam} as a function of the field strength $A$ for $\Omega = 10.5$ and $U=3$. The inset shows a field range for both frequencies $\Omega = 10.5$ and $\Omega=21$ in which the nonlocal doublon interaction $V$ changes signs, whereas the doublon hopping amplitude $J$ does not.}
\end{figure}

Single-particle hopping processes that change the number of doubly-occupied sites also change the total energy of the system. Here, we focus on the low-energy physics of doublons and disregard such processes in the Schrieffer-Wolff transformation of Hamiltonian (\ref{eq:Hp}). This leads to an effective Hamiltonian that describes the doublon subsystem in the nonequilibrium steady-state
\begin{align}
H^{d}(A) = \sum_{\av{ij}}J(A)\,d^{\dagger}_{i}d^{\phantom{\dagger}}_{j}
- \sum_{\av{ij}} V(A)\,\rho^{\phantom{\dagger}}_{i}\rho^{\phantom{\dagger}}_{j},
\label{eq:EffHam}
\end{align}
where $\rho^{\phantom{\dagger}}_{i} = d^{\dagger}_{i}d^{\phantom{\dagger}}_{i} = n^{\phantom{\dagger}}_{i\uparrow} n^{\phantom{\dagger}}_{i\downarrow}$ is the local double occupancy operator.
The hopping amplitude $J(A)$ and nonlocal interaction potential $V(A)$ of doublons depend on the strength of the external field as~\cite{SM}
\begin{align}
J(A) &= \frac{2t^2}{U} \left( {\cal J}^2_{0}(A)  - \frac{2U^2}{\Omega^2}\sum_{m>0}
\frac{(-1)^{m}}{m^2}{\cal J}^2_{m}(A)  \right),  
\label{eq:Td}\\
V(A) &= \frac{2t^2}{U} \left({\cal J}^2_{0}(A) - \frac{2U^2}{\Omega^2}\sum_{m>0}
\frac{1}{m^2}{\cal J}^2_{m}(A) \right),
\label{eq:Vd}
\end{align}
where ${\cal J}_{m}(A)$ is the $m$-th order Bessel function of the first kind. 
The zeroth order contribution in the limit $U\ll\Omega$ renormalizes the doublon hopping and interaction in the same way $J(A) = V(A) = {\cal J}^2_{0}(A)\,2t^2/U+O(U/\Omega)$.
As noticed in Ref.~\cite{PhysRevB.94.174503}, the field-dependent factor ${\cal J}^2_{0}(A)$ simply acts as an overall scaling parameter and results in trivial physics.
Thus, the doublon hopping and interaction both remain positive, as in equilibrium.
Here we go further and consider the second-order contributions in the high-frequency limit~\cite{SM}.
We find that these contributions of order $U^2/\Omega^2$ in Eqs.~\eqref{eq:Td} and~\eqref{eq:Vd} now lift the degeneracy between $J$ and $V$.
In particular, this effect becomes important for field amplitudes near $A\simeq2.4$, where the zeroth-order contribution vanishes.
Figure~\ref{fig:TV} shows that this even allows the doublon interaction $V$ to become repulsive out of equilibrium, while the hopping amplitude $J$ does not change sign.
This shows that an independent control of $V$ and $J$ is not only possible in the vicinity of the resonance $\Omega=U$~\cite{PhysRevB.94.174503}, but also in the high-frequency limit (where energy absorption is well controlled).

\begin{figure}[t!]
\includegraphics[width=0.95\linewidth]{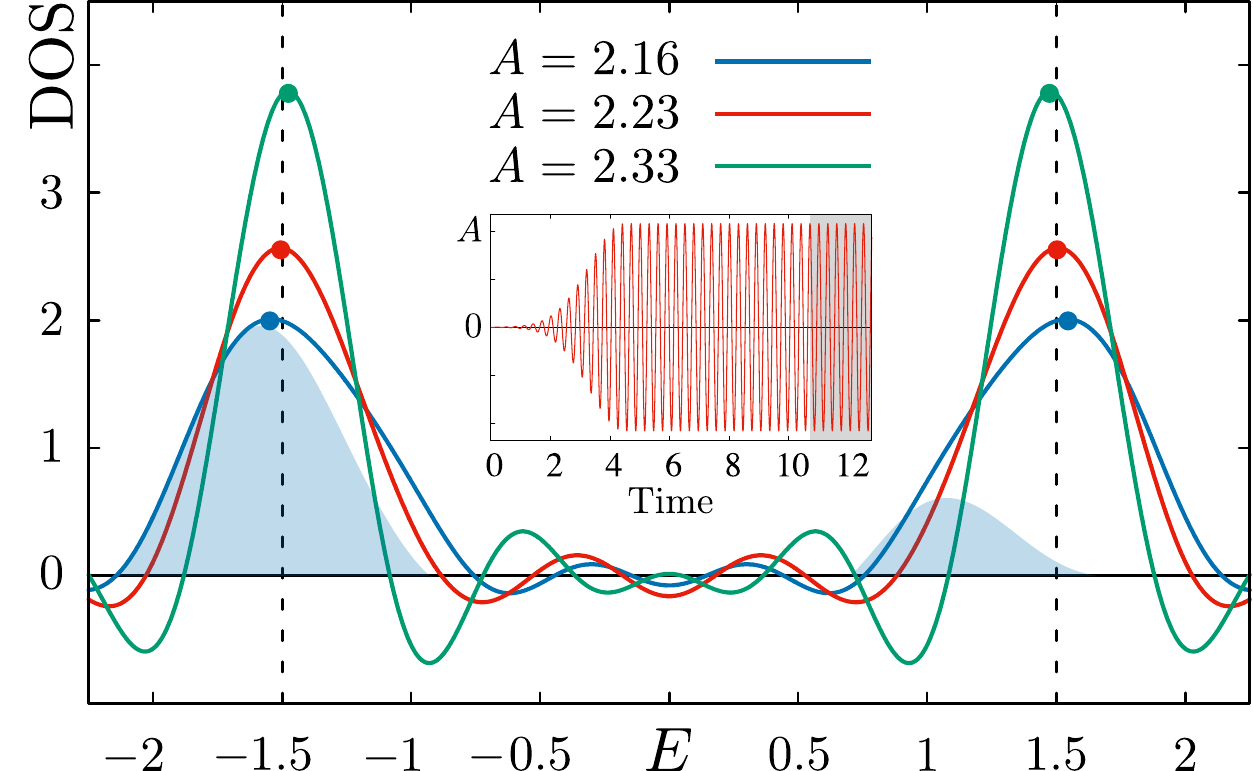}
\caption{\label{fig:dos} Time-averaged fermionic DOS obtained for $\Omega = 21$ and various field strengths $A=2.16$ (blue), $A=2.23$ (red), and ${A=2.33}$ (green). Colored points mark the top of the Hubbard sub-bands. Vertical dashed lines at $E=\pm{}U/2=\pm1.5$ is a guide to the eyes. Shaded blue areas indicate the time-averaged population of the upper and lower sub-bands at the field $A=2.16$. The inset shows the time-profile of the vector potential. The time-averaging is performed over the last 8 periods highlighted in grey.}
\end{figure}

\begin{figure}[t!]
\includegraphics[width=0.95\linewidth]{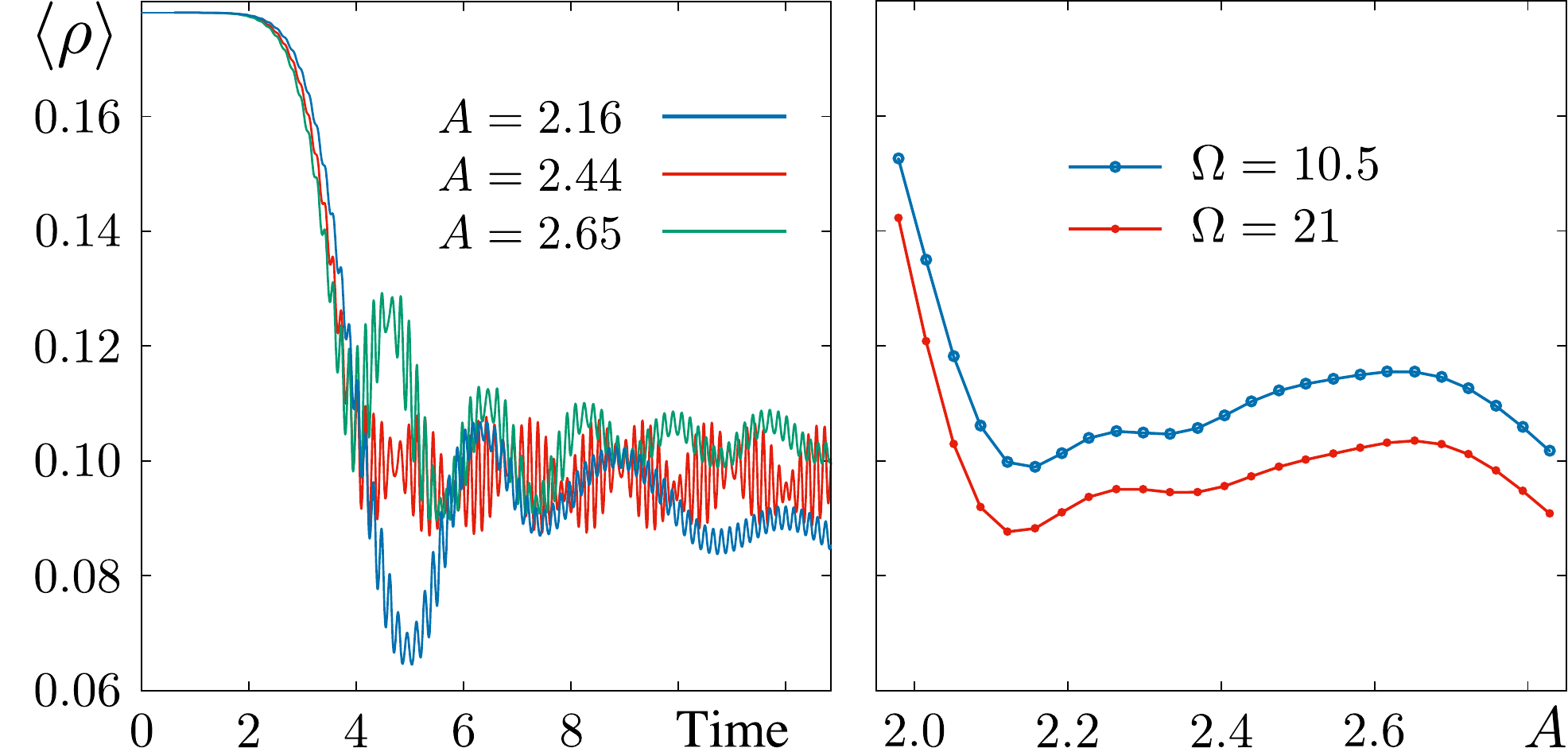}
\caption{\label{fig:Docc} (Left panel) Doublon density as a function of time obtained for three different values of the vector potential $A=2.16$ (blue line), $A=2.44$ (red line), and $A=2.65$ (green line). (Right panel) Time-averaged doublon density as a function of vector potential obtained for two frequencies $\Omega = 10.5$ (blue line) and $\Omega=21$ (red line).}
\end{figure}

{\it Nonequilibrium Hubbard sub-bands ---}
The switch of the doublon interaction may in principle be observed in experiment through a measurement of the doublon-doublon susceptibility.
However, the latter is hardly accessible, as it corresponds to a four-fermion response function.
Instead, we now show that signatures of the doublon interaction can also be revealed through the single-fermion density of states (DOS), as routinely measured in experiments~\cite{Neddermeyer_1,COURTHS198453,RevModPhys.75.473}.
To illustrate this point, we perform time-dependent DMFT simulations for the half-filled Hubbard model~\eqref{eq:Hamiltonian1}~\cite{RevModPhys.86.779}.
We consider the local Coulomb interaction $U=3$, so that the driven system for the range of fields $A$ when the doublon interaction $V(A)$ changes sign lies in a Mott-insulating state. Values $\Omega=10.5$ and $\Omega=21$ for the frequency of the field are taken to justify the requirement $(U/\Omega)^2\ll1$ for the high-frequency expansion. The energy is given in units of the electron hopping amplitude $t$.
The ac field is turned on up to the maximum value $A_{\rm max} = \sqrt{2}A$ following the exponential ramp $\exp \left\{-(t-t_0)^2/(2\sigma^2)\right\}$ in order to avoid heating~\cite{PhysRevLett.106.236401, 2016arXiv161105024H}. Here $\sigma=\frac{d}{2\sqrt{2ln2}}$ and $d$ is a full width at half-maximum of a pulse. The time profile of the field is shown in the inset of Fig.~\ref{fig:dos}. 
We then determine the fermion DOS given by the spectral function $A^{R}(t, E)$ defined as (see, e.g. Ref.~\onlinecite{PhysRevLett.107.186406})
\begin{align}
\label{eq:Spectr_R_L}
A^{\alpha}(t, E) =-\dfrac{1}{\pi} {\rm Im} \int_{0}^{s_{\rm max}} ds \, e^{iE s} G^{\alpha}_{\rm loc}(t,t-s),
\end{align}
where for $G^{\alpha}_{\rm loc}(t,t')$ we take the local retarded Green's function $G^{R}_{\rm loc}(t,t')=-i \theta(t-t') \, \langle \{c^{\phantom{!}}_{i}(t),  c^{\dagger}_{i}(t')\} \rangle$.
We perform the numerical time-dependent DMFT calculations within the iterative perturbation theory (IPT)~\cite{PhysRevLett.107.186406} on a $32\times32$ $k$-grid starting with the inverse temperature $\beta=5$.
We finally average over the last 8 periods of time (gray area in the inset of Fig.~\ref{fig:dos}), for which the behavior of time-resolved observables, such as the double occupancy presented in Fig.~\ref{fig:Docc} (left),
indicates that the system is in a nonequilibrium steady state.
This will allow us to compare the numerical simulations with the effective time-independent description in Eq.~\eqref{eq:Hp}.

Fig.~\ref{fig:dos} shows the single-fermion DOS resulting from the nonequilibrium DMFT simulations for $\Omega=21$ and various field amplitudes. 
The two spectral peaks below and above the Fermi energy ($E=0$) are the lower and upper Hubbard sub-bands that correspond to doublon and holon (fully unoccupied site) states, respectively.
This structure of the DOS confirms that, for the range of field amplitudes we consider, the interacting fermion system is indeed a Mott-insulator.
An occupation function can be obtained as a time-averaged spectral function $A^{<}(t, E)$~\eqref{eq:Spectr_R_L} of the lesser Green's function $G^{<}_{\rm loc}(t,t') = i \langle c^{\dagger}_{i}(t') c^{\phantom{!}}_{i}(t) \rangle$.  Shaded blue areas in Fig.~\ref{fig:dos} show that the upper Hubbard sub-band is only slightly populated upon driving. 
An effective temperature of the system can be estimated from the nonequilibrium distribution function $A^{<}(t, E)/A^{R}(t, E)$, and gives an effective inverse temperature
of order $\beta=2$.

\begin{figure}[t!]
\includegraphics[width=0.95\linewidth]{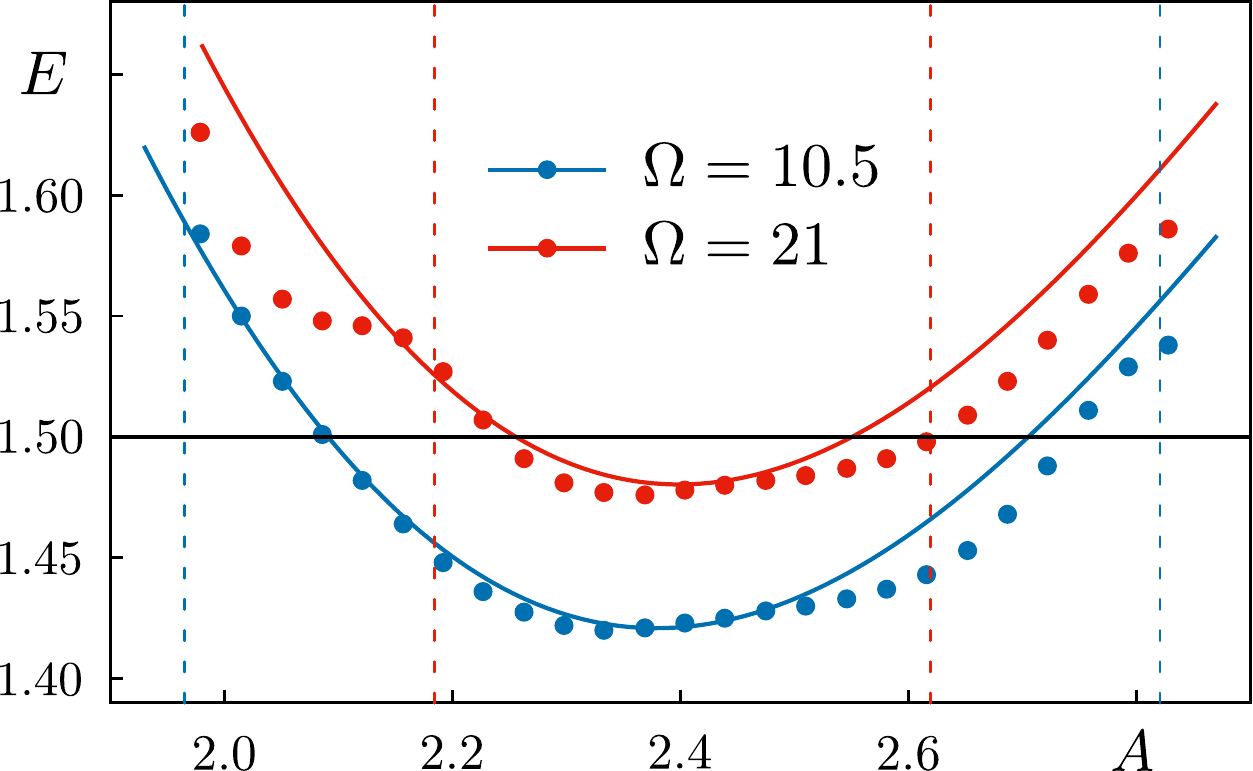}
\caption{\label{fig:UHB} Position of the top of the upper Hubbard sub-band as a function of the applied light $A$. Results are obtained for two frequencies $\Omega = 10.5$ (blue color) and $\Omega=21$ (red color). Solid lines correspond to the estimation $U_{\rm eff}/2$ obtained from the effective local Coulomb interaction. Points correspond to the nonequilibrium DMFT result. Vertical dashed lines indicate fields at which the doublon-doublon interaction $V(A)$ changes sign. Horizontal line $E=U/2=1.5$ serves as a guide to the eyes. }
\end{figure}

We find that, varying the field amplitude results in the energy shift of the Hubbard sub-bands.
It is not surprising, since the position of the sub-bands is determined by the local Coulomb interaction as $E\simeq\pm U'(A)/2$, which explicitly depends on the field amplitude.
Points in Fig.~\ref{fig:UHB} represent the peak energy of the upper Hubbard sub-band as a function of $A$ obtained from nonequilibrium DMFT simulations for two different frequencies.
While $A$ increases, we observe that the sub-bands first move closer to the Fermi level as if they attract each other. Above a critical field, the interaction between peaks switches to repulsive, and the distance between them increases again.
As we now show, this behavior of Hubbard sub-bands is a manifestation of an attractive-repulsive transition of the interaction between doublons.

{\it Detection of repulsive doublon interactions ---}
We find that the local Coulomb potential $U'(A)$, alone, cannot explain the behavior of Hubbard sub-bands obtained from DMFT simulations in Fig.~\ref{fig:UHB}. 
As shown in SM~\cite{SM}, $U'(A)<U$ for any field amplitude $A$.
Then, the peak of the upper Hubbard sub-band, if given by $U'(A)/2$, cannot exceed the equilibrium value of $U/2$.
For this reason, we further account for the effects of the non-local interactions and hopping processes in Hamiltonian~\eqref{eq:Hp} on the position of the sub-bands.
We first map the nonlocal Coulomb potential onto an on-site potential through the Peierls-Feynman-Bogoliubov variational principle~\cite{PhysRev.54.918, bogolyubov1958variation, feynman1972statistical}.
This leads to the local Coulomb interaction ${U^{*}(A) = U'(A) - V'(A)}$~\cite{PhysRevLett.111.036601}.
The effect of other terms in Hamiltonian~\eqref{eq:Hp} can be taken into account perturbatively. In particular, we find that only the two-hopping processes contribute to the effective local Coulomb potential $U_{\rm eff}$~\cite{SM}.
This finally results in
\begin{align}
U_{\rm eff}(A) &= U + \frac{2t^2}{U}\left(\frac{1}{\av{\rho}}{\cal J}^2_{0}(A) - \frac{9U^2}{\Omega^2}\sum_{m>0}
\frac{1}{m^2}{\cal J}^2_{m}(A) \right). 
\label{eq:Ueff}
\end{align}
Thus, the behavior of Hubbard sub-bands as the function of field can be approximated by the following relation ${E=U_{\rm eff}(A)/2}$. The mean value of the double occupancy $\av{\rho}$ that enters the Eq.~\eqref{eq:Ueff} can be extracted from Fig.~\ref{fig:Docc}, which represents the nonequilibrium DMFT result. For two different frequencies $\Omega=10.5$ and $\Omega=21$ we find $\av{\rho}\simeq0.11$ and $\av{\rho}\simeq0.10$, respectively.
Remarkably, solid lines in Fig.~\ref{fig:UHB} show that this simple estimation for the position of the sub-bands based on the effective time-independent description of the problem~\eqref{eq:Hp} accurately reproduces the result of nonequilibrium DMFT simulations for the initial time-dependent model~\eqref{eq:Hamiltonian1}. This fact suggests that the introduced Hamiltonian~\eqref{eq:Hp} correctly describes properties of the nonequilibrium steady state of the system. In addition, the relation~\eqref{eq:Ueff} can also serve as the measure for the average double occupancy of the lattice site $\av{\rho}$.

We can further relate the effective local Coulomb potential $U_{\rm eff}$ to the doublon interaction $V(A)$ in Hamiltonian~\eqref{eq:EffHam}.
In equilibrium, the mean double occupancy $\av{\rho}$ per lattice site does not exceed $1/4$ at half-filling.
Larger values of $\av{\rho}$ indicate a population inversion.
This can occur out-of equilibrium but for much stronger fields and smaller frequencies of the field ($\Omega\sim{}U$ close to a resonant driving between Hubbard sub-bands) than the ones we are dealing with here~\cite{PhysRevLett.105.220405, PhysRevB.101.161101, Herrmann_2017}.
In our case, Fig.~\ref{fig:Docc} shows that the double occupancy does not exceed $1/4$.
Thus, the attraction-repulsion transition of doublon interactions in Fig.~\ref{fig:TV} does not involve any population inversion, in contrast to previous proposal~\cite{PhysRevB.85.155124}.
Besides, if we consider the maximum value of the mean double occupancy, i.e. $\av{\rho}=1/4$, the energy shift of the Hubbard sub-bands with respect to their equilibrium position in the atomic limit ($\pm U/2$) is
\begin{align}
\Delta(A) = \left[U_{\rm eff}(A) - U\right]/2 \simeq 2 V(A).   
\end{align}
Therefore, this shift $\Delta$ is a single-fermion measurement of the strength of the doublon-doublon interaction $V(A)$ for ${\av{\rho}=1/4}$.
For smaller values of $\av{\rho}$, we more generally find ${\Delta(A)\ge2V(A)}$.
It follows that the negative value of ${\Delta(A)<0}$ obtained in the nonequilibrium DMFT simulations in Fig.~\ref{fig:UHB} is an indirect indication that the doublon interaction, initially attractive at zero field, has become repulsive out of equilibrium.
The value of the field at which the interaction between doublons changes sign is depicted in Fig.~\ref{fig:UHB} by vertical dashed lines. 
Importantly, the total suppression of the doublon interaction ${V(A)=0}$ happens at a considerably smaller power of the field $A$ compared to the regime of dynamical localization, determined by the first root of the Bessel funtion ${{\cal J}_0(A)=0}$. Therefore, the high-frequency driving provides a unique possibility to explore the regime of a weakly interacting doublon liquid, whereas in equilibrium doublons are strongly interacting $J(0)=V(0)$ as follows from Eqs.~\eqref{eq:Td} and~\eqref{eq:Vd}.

{\it Conclusions ---}
To conclude, in this work we have studied the effect of the applied high-frequency perturbation on the doublon subsystem of the fermion Hubbard model. First, we have introduced an effective time-independent Hamiltonian that describes a time-averaged dynamics of doublons. We have shown that the hopping amplitude and nonlical interaction of this effective model can be controlled by the value of the vector potential. Moreover, in a certain range of fields, a possibility for a dynamical attraction-repulsion transition of doublons has been investigated. Further, we have shown that the signature of this transition can be found in the behavior of Hubbard sub-bands of the single-fermion density of states. We have proposed a simple explanation of the observed effect based on the renormalization of the local Coulomb potential via hopping processes and nonlocal Coulomb interaction. The obtained result provides a clear criterion for the experimental confirmation of the repulsive interactions between doublons that involves only a local single-fermion observable.\\

\begin{acknowledgments}
The authors thank Michael Potthoff and Sergey Brener for useful discussions and comments. The authors acknowledge the support by the Cluster of Excellence ``Advanced Imaging of Matter'' of the Deutsche Forschungsgemeinschaft (DFG) - EXC 2056 - Project No. ID390715994. This research was also supported through the European XFEL and DESY computational resources in the Maxwell infrastructure operated at Deutsches Elektronen-Synchrotron (DESY), Hamburg, Germany. C.D. acknowledges the support of Idex Bordeaux (Maesim Risky project 2019 of the LAPHIA Program). The work of V.N.V., M.I.K. and A.I.L. is supported by European Research Council via Synergy
Grant 854843 - FASTCORR.
\end{acknowledgments}

\bibliography{Ref}

\clearpage
\appendix

\onecolumngrid
\begin{center}
{\bf \large Supplemental material}\\[1cm]
\end{center}
\twocolumngrid

\section*{Effective time-independent Hamiltonian}
In this section we derive an effective time-independent Hamiltonian that describes the Hubbard model under the high-frequency periodic driving.
Let us start with the following Hamiltonian of the Hubbard model 
\begin{align}
H &= \sum_{\av{ij},\,\sigma}t^{\phantom{*}}_{ij}(\tau)c^{*}_{i\sigma}c^{\phantom{*}}_{i\sigma}
+\sum_{i}U
n^{\phantom{*}}_{i\uparrow}n^{\phantom{*}}_{i\downarrow},
\label{eq:Hamiltonian1_app}
\end{align}
where $\tau=\Omega{}t$, $t_{ij}(\tau) = t e^{-i{\bf A}(\tau)\cdot{\bf R}_{ij}}$, and ${\bf R}_{ij}$ is a unit vector that connects neighbouring sites $\av{ij}$. All notations are introduced in the main text of the paper. Here we take the vector potential ${\bf A}(\tau)= {\bf A} \cos(\tau)$ in the temporal gauge. 

After we accounted for the high-frequency perturbation via the Peierls substitution, the Hamiltonian~\eqref{eq:Hamiltonian1_app} becomes time periodic. Its quantum nonequilibrium steady states obey the time-dependent Schr\"odinger equation 
\begin{align}
i \partial_\tau \Psi(\lambda, \tau) =~\frac{1}{\Omega} H(\tau) \Psi (\lambda, \tau).
\end{align}
One can introduce a dimensionless parameter $\lambda=\delta E/ \Omega$ which compares a certain energy scale $\delta E$ to the field frequency. For simplicity we chose $\delta E=U$ as the largest characteristic energy involved in Hamiltonian~\eqref{eq:Hamiltonian1_app}.
Then, the Schr\"odinger equation reads
\begin{align}
i \partial_\tau \Psi(\lambda, \tau) =~\lambda\overline{H}(\tau) \Psi (\lambda, \tau),
\end{align} 
where the Hamiltonian is now rescaled on the same energy as $\overline{H}(\tau)=H(\tau)/\delta{}E$. In the high-frequency limit, $\lambda$ is a small parameter and we can look for a unitary transformation defined as $\Psi (\lambda, \tau) =~\exp\{-i\Delta(\tau)\}\,\psi (\lambda, \tau)$ which removes the time dependence of the Hamiltonian~\cite{ITIN2014822, PhysRevLett.115.075301, PhysRevB.93.241404, PhysRevB.95.024306, PhysRevLett.118.157201}. By construction we also impose $\Delta(\tau) =~ \sum_{n=1}^{+\infty} \lambda^{n} \Delta_{n}(\tau)$, with 
$\Delta_{n}(\tau)$ a $2\pi$ periodic function. Such a transformation leads to 
\begin{align}
i\partial_\tau \psi(\lambda, \tau) = \frac{1}{\Omega}{\cal H} \psi (\lambda, \tau) = \lambda \overline{\cal H} \psi (\lambda, \tau),
\end{align}
with effective Hamiltonian
\begin{align}\label{Effective Hamiltonian}
\overline{\cal H}=e^{i\Delta(\tau)} \overline{H}(\tau)\,e^{-i\Delta(\tau)} -i\lambda^{-1} e^{i\Delta(\tau)} \partial_{\tau} e^{-i\Delta(\tau)} ~,
\end{align}
or equivalently
\begin{align}
{\cal H}=e^{i\Delta(\tau)}H(\tau)\,e^{-i\Delta(\tau)} -i\Omega e^{i\Delta(\tau)} \partial_{\tau} e^{-i\Delta(\tau)} ~.
\end{align}
The partial time-derivative satisfies the following relation
\begin{align}\label{Exponential Operator Derivative}
\partial_{\tau} e^{-i\Delta(\tau)} &= \sum_{n=0}^{\infty} \frac{ \big\{ \big(\!-i\Delta(\tau) \big)^{n}, -\,i \partial_{\tau}\Delta(\tau) \big\}}{(n+1)!}~ e^{-i\Delta(\tau)} ~,
\end{align}
where the repeated commutator is defined for two operators $X$ and $Y$ by $\{1, Y\}=Y$ and $\{X^{n}, Y\}=~[X, \{ X^{n-1}, Y\}]$. The square brackets denote the usual commutator. Then, one can write
\begin{align}
\overline{\cal H}= e^{i\Delta(\tau)} \left[\overline{H}(\tau) -i\lambda^{-1} \sum_{n=0}^{\infty} \frac{ \big\{ \big(\!-i\Delta(\tau) \big)^{n}, -\,i \partial_{\tau}\Delta(\tau) \big\}}{(n+1)!} \right] e^{-i\Delta(\tau)} ~,
\end{align}
Using the series representation
\begin{align} \overline{\cal H}=\sum_{n=0}^{\infty}\lambda^{n}
\tilde{H}_{n},
\end{align}
together with Eqs. \eqref{Exponential Operator Derivative} and \eqref{Effective Hamiltonian}, one can then determine operators $\tilde{H}_{n}$ and $\Delta_{n}$ iteratively in all orders in $\lambda$. Here, we restrict ourselves to the case of the high-frequency field, which allows us to consider the effective Hamiltonian representation up to the second order correction in $\lambda$: $\overline{\cal H} = \tilde{H}_{0} + \lambda\tilde{H}_{1} + \lambda^{2}\tilde{H}_{2}$. These effective time-independent Hamiltonians describe the stroboscopic dynamics of the system, whereas its evolution between two stroboscopic times is encoded into the time-dependent function $\Delta_{n}(\tau)$.
For the considered problem~\eqref{eq:Hamiltonian1_app}, the effective time-independent Hamiltonian 
was explicitly obtained in the Ref.~\onlinecite{PhysRevLett.118.157201} and reads
\begin{align}
H'(A) &= \sum_{\av{ij},\,\sigma}t'(A)c^{\dagger}_{i\sigma}c^{\phantom{\dagger}}_{j\sigma}
+ U'(A) \sum_{i} n^{\phantom{\dagger}}_{i\uparrow}n^{\phantom{\dagger}}_{i\downarrow} \notag\\
&+ \sum_{\av{ij}} \left(J'(A)\,d^{\dagger}_{i}d^{\phantom{*}}_{j} + \frac12V'(A)\,n_{i}n_{j}
+ \frac12{\cal I}'(A)\,{\bf S}_{i}^{\phantom{*}}{\bf S}_{j}^{\phantom{*}}\right).
\label{eq:Hp_app}
\end{align}
Here, the hopping amplitude
\begin{align}
t'(A) = t\,{\cal J}_0(A)
\end{align}
is renormalized by the Bessel function of the zeroth order ${\cal J}_{0}(A)$. The local and nonlocal Coulomb interactions are
\begin{align}
U'(A) &= U - \frac{16t^2U}{\Omega^2}\sum_{m>0}
\frac{1}{m^2}{\cal J}^2_{m}(A), \\
V'(A) &=  \frac{2t^2U}{\Omega^2}\sum_{m>0}
\frac{1}{m^2}{\cal J}^2_{m}(A),
\label{eq:Vp_app}
\end{align}
and the field-induced hopping process of doublons $d_{j} = c_{j\downarrow}c_{j\uparrow}$ has the following amplitude
\begin{align}
J'(A) &= -  \frac{4t^2U}{\Omega^2}\sum_{m>0}
\frac{(-1)^{m}}{m^2}{\cal J}^2_{m}(A).
\label{eq:Jp_app}
\end{align}
The behavior of the exchange interaction ${\cal I}'(A)$ has been discussed in details in the Ref.~\onlinecite{PhysRevLett.118.157201}, and is not of the interest for the current work.

\section*{Effective local Coulomb potential}

An effective local Coulomb potential $U_{\rm eff}$ for the Hamiltonian~\eqref{eq:Hp_app} that determines the position of Hubbard sub-bands in the single-fermion density of states (DOS) can be obtained in the framework of the action formalism. For simplicity, the result is obtained perturbatively at a finite temperature $T$ that will be set to zero at the end of the calculation.
The action that corresponds to the Hamiltonian~\eqref{eq:Hp_app} is following
\begin{align}
{\cal S} = &-\frac{1}{\beta}\sum_{\av{ij},\nu,\sigma} c^{*}_{i\nu\sigma} \left[i\nu + \mu - t'(A)\right] c^{\phantom{*}}_{j\nu\sigma} + U^{*}(A) \sum_{i} n^{\phantom{\dagger}}_{i\uparrow}n^{\phantom{\dagger}}_{i\downarrow} \notag\\
&+ \sum_{\av{ij}} \left(J'(A)\,d^{\dagger}_{i}d^{\phantom{*}}_{j}
+ \frac12{\cal I}'(A)\,{\bf S}_{i}^{\phantom{*}}{\bf S}_{j}^{\phantom{*}}\right),
\label{eq:S1}
\end{align}
where we also accounted for the effect of the nonlocal Coulomb interaction via the Peierls-Feynman-Bogoliubov variational principle, which gives ${U^{*}(A) = U'(A) - V'(A)}$ (see the main text). Here, $\beta=1/T$ is the inverse temperature. 
We aim to map the introduced action onto a purely local problem
\begin{align}
{\cal S}_2 = &-\frac{1}{\beta}\sum_{i,\nu,\sigma} c^{*}_{i\nu\sigma} \left[i\nu + \mu \right] c^{\phantom{*}}_{i\nu\sigma} + U_{\rm eff}(A)\sum_{i}n_{i\uparrow}n_{i\downarrow}
\label{eq:S2}
\end{align}
with $U_{\rm eff}(A) = U^{*}(A) + \delta{}U(A)$ that determines the position of Hubbard sub-bands as $E=\pm{}U_{\rm eff}(A)/2$. Now, we can expand the partition function ${\cal Z}$ of the action~\eqref{eq:S1} up to the second order in the small parameter $t'/U$, and compare the contribution to the total energy with the first order expansion of the partition function for the action~\eqref{eq:S2} in terms of $\delta{}U(A)$. 
This results in
\begin{widetext}
\begin{align}
{\cal Z}_1 &= \int D[c^{*},c] \, e^{-\beta {\cal S}_{1}} 
\notag\\
&= \int D[c^{*},c] \, e^{-\beta {\cal S}_{0}}
\Big[ 1 - \sum_{\av{ij},\nu,\sigma} \left(t'(A)\av{c^{*}_{i\nu\sigma} c^{\phantom{*}}_{j\nu\sigma}} 
+ \beta J'(A) \av{d^{\dagger}_{i}d^{\phantom{*}}_{j}}
+ \frac{\beta{\cal I}'(A)}{2}\av{{\bf S}_{i}^{\phantom{*}}{\bf S}_{j}^{\phantom{*}}}\right) + \frac{t'^2}{2} \sum_{\substack{\av{ij},\nu,\sigma \\ \av{kl},\nu',\sigma'}} \av{c^{*}_{i\nu\sigma} c^{\phantom{*}}_{j\nu\sigma}  c^{*}_{k\nu'\sigma'} c^{\phantom{*}}_{l\nu'\sigma'}} \Big],
\label{eq:Z1}
\end{align}
\end{widetext}
where the action
\begin{align}
{\cal S}_0 = &-\frac{1}{\beta}\sum_{i,\nu,\sigma} c^{*}_{i\nu\sigma} \left[i\nu + \mu \right] c^{\phantom{*}}_{i\nu\sigma} + U^{*}(A)\sum_{i}n_{i\uparrow}n_{i\downarrow},
\label{eq:S0}
\end{align}
is purely local. Thus, all terms Eq.~\eqref{eq:Z1} contained in round brackets are zero. The last term can be simplified as
\begin{align}
\frac{t'^2}{2} \sum_{\substack{\av{ij},\nu,\sigma \\ \av{kl},\nu',\sigma'}} \av{c^{*}_{i\nu\sigma} c^{\phantom{*}}_{j\nu\sigma}  c^{*}_{k\nu'\sigma'} c^{\phantom{*}}_{l\nu'\sigma'}} &= 
-t'^2\sum_{\av{ij},\nu} \av{c^{*}_{i\nu\uparrow} c^{\phantom{*}}_{i\nu\uparrow}} \av{c^{*}_{j\nu\uparrow} c^{\phantom{*}}_{j\nu\uparrow}} \notag\\
&= -4 t'^{2} \sum_{i,\nu} g^{2}_{i\nu\uparrow},
\end{align}
where the coefficient ``4'' corresponds to a number of nearest-neighbor lattice sites on a 2D square lattice. $g_{i\nu\sigma}$ is the exact Green's function of the local problem~\eqref{eq:S0}, which can be found from the following relation (see, e.g.~\cite{ayral2015nonlocal})
\begin{align}
g^{-1}_{i\nu\sigma} = i\nu - \frac{U^{*\,2}(A)}{4i\nu}.
\end{align}
Taking the zero limit for the temperature ($\beta\to\infty$), the sum over Matsubara frequencies $\nu_{n}=\pi(2n+1)/\beta$ can be replaced by the integral, and we get 
\begin{align}
-4t'^2\sum_{i,\nu} g^2_{i\nu} &= 
-4t'^2\sum_{i,\nu} \frac{\nu^2}{\left(\nu^2 + U^{*\,2}(A)/4\right)^2} \notag\\
&= -\frac{t'^2\beta^2}{\pi^2} \sum_{i} \int_{-\infty}^{+\infty} \frac{x^2 dx}{\left(x^2 + a^2\right)^2} \notag\\
&= - \sum_{i}\frac{t'^2\beta^2}{2\pi{}a} \notag\\
&\simeq -\beta \sum_{i}\frac{2t'^2}{U},
\end{align}
where $a=\beta{}U^{*}(A)/(4\pi)$.

For the second partition function we get
\begin{align}
{\cal Z}_2 &= \int D[c^{*},c] \, e^{-\beta {\cal S}_{2}} 
\notag\\
&= \int D[c^{*},c] \, e^{-\beta {\cal S}_{0}}
\Big[ 1 - \beta\sum_{i} \delta{}U \av{n_{i\uparrow}n_{i\downarrow}}  \Big].
\label{eq:Z1}
\end{align}
Thus, the correction to the effective local Coulomb potential can be identified from the relation $\delta{}U = 2t'^{2}/(U \av{\rho})$, where $\av{\rho}$ is the mean double occupancy of the system. Therefore, the total effective local Coulomb interaction reads
\begin{align}
U_{\rm eff}(A) &= U + \frac{2t^2}{U}\left(\frac{1}{\av{\rho}}{\cal J}^2_{0}(A) - \frac{9U^2}{\Omega^2}\sum_{m>0}
\frac{1}{m^2}{\cal J}^2_{m}(A) \right). 
\label{eq:Ueff_app}
\end{align}

\section*{Doublon Hamiltonian}

An effective doublon Hamiltonian can be obtained performing the Schrieffer-Wolff transformation~\cite{CHAO1977163, Chao_1977, PhysRevB.37.9753, spalek2007tj} of the fermion problem~\eqref{eq:Hp_app}. This transformation excludes single electron hopping processes that change the number of doubly occupied sites in the system. To identify them, the kinetic part of the Hamiltonian~\eqref{eq:Hp_app} can be multiplied by the unity $1=n_{i,\sigma} + h_{i,\sigma}$, where $n_{i,\sigma}$ is an electronic density with spin $\sigma$ on a site $i$, and $h_{i,\sigma}=1-n_{i,\sigma}$ is a hole density on the same site with the same spin. This results in
\begin{align}
\sum_{\av{ij},\,\sigma}t'_{ij}c^{*}_{i\sigma}c^{\phantom{*}}_{j\sigma} = 
{\cal H}_{t}^{0} + {\cal H}_{t}^{+} + {\cal H}_{t}^{-},
\end{align}
where the first contribution
\begin{align}
{\cal H}_{t}^{0} &= \sum_{\av{ij},\,\sigma}t'_{ij}n_{i,-\sigma}c^{*}_{i\sigma}c^{\phantom{*}}_{j\sigma}n_{j,-\sigma} +
\sum_{\av{ij},\,\sigma}t'_{ij}h_{i,-\sigma}c^{*}_{i\sigma}c^{\phantom{*}}_{j\sigma}h_{j,-\sigma}
\end{align}
does not change the number of the doubly-occupied sites. Other two 
\begin{align}
{\cal H}_{t}^{+} + {\cal H}_{t}^{-} = \sum_{\av{ij},\,\sigma} \left(t'_{ij}n_{i,-\sigma}c^{*}_{i\sigma}c^{\phantom{*}}_{j\sigma}h_{j,-\sigma} + 
t'_{ij}h_{i,-\sigma}c^{*}_{i\sigma}c^{\phantom{*}}_{j\sigma}n_{j,-\sigma} \right)
\end{align}
increase (decrease) this number by one, respectively. The term ${\cal H}_{t}^{0}$ is not relevant for the Mott-insulating regime and can be neglected. Last two contributions ${\cal H}_{t}^{+} + {\cal H}_{t}^{-}$ can be eliminated introducing a proper unitary transformation~\cite{CHAO1977163, Chao_1977, PhysRevB.37.9753, spalek2007tj}. Then, one can obtain an effective Hamiltonian that describes the doublon subsystem up to the first order in the small parameter $t'/U$
\begin{align}
{\cal H} = \sum_{\av{ij}}J(A)\,d^{\dagger}_{i}d^{\phantom{\dagger}}_{j}
+ \sum_{i}\bar{U}(A)\,n^{\phantom{*}}_{i\uparrow}n^{\phantom{*}}_{i\downarrow}
+ \frac12 \sum_{\av{ij}}\bar{V}(A)\, n^{\phantom{*}}_{i}n^{\phantom{*}}_{j}.
\label{eq:EffHam_app}
\end{align}
The second term can be excluded from the Hamiltonian, because it plays the role of the chemical potential for doublons, since $\rho_{i} = n^{\phantom{*}}_{i\uparrow}n^{\phantom{*}}_{i\downarrow}$. The last term can also be transformed as
\begin{align}
n^{\phantom{*}}_{i,\sigma}n^{\phantom{*}}_{j,\sigma'} 
&= 
(n^{\phantom{*}}_{i,-\sigma}+h^{\phantom{*}}_{i,-\sigma})
n^{\phantom{*}}_{i,\sigma}n^{\phantom{*}}_{j,\sigma'}
(n^{\phantom{*}}_{j,-\sigma'}+h^{\phantom{*}}_{j,-\sigma'}) \notag\\
&\simeq 
n_{i,-\sigma}n_{i,\sigma}n_{j,\sigma'}
n_{j,-\sigma'} = 4\,\rho_{i}\,\rho_{j}
\end{align}
leading to the nonlocal interaction between doublons. In the last equation, terms that contain $h^{\phantom{*}}_{i,-\sigma}n^{\phantom{*}}_{i,\sigma}$ have also been nenglected, because they are zero in the doubly occupied case. This results in the final Hamiltonian for doublon degrees of freedom 
\begin{align}
H^{d}(A) = \sum_{\av{ij}}J(A)\,d^{\dagger}_{i}d^{\phantom{\dagger}}_{j}
- \sum_{\av{ij}} V(A)\,\rho^{\phantom{\dagger}}_{i}\rho^{\phantom{\dagger}}_{j},
\label{eq:EffHam2_app}
\end{align}
where the hopping amplitude $J(A)$ and nonlocal interaction $V(A)$ of doublons are
\begin{align}
J(A) &= \frac{2t^2}{U} \left( {\cal J}^2_{0}(A)  - \frac{2U^2}{\Omega^2}\sum_{m>0}
\frac{(-1)^{m}}{m^2}{\cal J}^2_{m}(A)  \right), 
\label{eq:Td_app}\\
V(A) &= \frac{2t^2}{U} \left({\cal J}^2_{0}(A) - \frac{2U^2}{\Omega^2}\sum_{m>0}
\frac{1}{m^2}{\cal J}^2_{m}(A) \right).
\label{eq:Vd_app}
\end{align}
Here, the contribution proportional to ${\cal J}_0^2$ appears as the result of the Schrieffer-Wolff transformation, and the second term proportional to $U^2/\Omega^2$ follows form Eqs.~\eqref{eq:Vp_app} and~\eqref{eq:Jp_app}.

\end{document}